\documentclass[sigconf]{acmart}

\usepackage{graphicx}
\usepackage{amsmath}
\usepackage{booktabs}
\usepackage{algorithm}
\usepackage{algorithmic}
\usepackage{amsfonts}
\usepackage{multirow}
\usepackage{makecell}
\usepackage{subfigure}
\usepackage{color}
\usepackage{bm}
\usepackage{epstopdf}
\usepackage{url}
\usepackage[cal=cm]{mathalfa}
\usepackage{balance}
\usepackage{threeparttable}
\usepackage{lipsum}
\usepackage{enumitem}
 \usepackage{pifont}
 \usepackage{tabularx}
 \usepackage{makecell}

\AtBeginDocument{%
  \providecommand\BibTeX{{%
    \normalfont B\kern-0.5em{\scshape i\kern-0.25em b}\kern-0.8em\TeX}}}

\renewcommand\footnotetextcopyrightpermission[1]{}

\begin{document}
\title{KuaiFormer: Transformer-Based Retrieval at Kuaishou}
\renewcommand{\shorttitle}{KuaiFormer}

\author{Chi Liu}
\affiliation{
  \institution{Kuaishou Technology, Beijing, China}
  \country{liuchi05@kuaishou.com}
}

\author{Jiangxia Cao}
\affiliation{
  \institution{Kuaishou Technology, Beijing, China}
  \country{caojiangxia@kuaishou.com}
}

\author{Rui Huang}
\affiliation{
  \institution{Kuaishou Technology, Beijing, China}
  \country{huangrui06@kuaishou.com}
}

\author{Kai Zheng}
\affiliation{
  \institution{Kuaishou Technology, Beijing, China}
  \country{zhengkai@kuaishou.com}
}

\author{Qiang Luo}
\affiliation{
  \institution{Kuaishou Technology, Beijing, China}
  \country{luoqiang@kuaishou.com}
}

\author{Kun Gai}
\affiliation{
  \institution{Unaffiliated, Beijing, China}
  \country{gai.kun@qq.com}
}

\author{Guorui Zhou}
\affiliation{
  \institution{Kuaishou Technology, Beijing, China}
  \country{zhouguorui@kuaishou.com}
}

\begin{abstract}

In large-scale content recommendation systems, retrieval serves as the initial stage in the pipeline, responsible for selecting thousands of candidate items from billions of options to pass on to ranking modules. 
Traditionally, the dominant retrieval method has been Embedding-Based Retrieval (EBR) using a Deep Neural Network (DNN) dual-tower structure. 
However, applying transformer in retrieval tasks has been the focus of recent research, though real-world industrial deployment still presents significant challenges.
In this paper, we introduce KuaiFormer, a novel transformer-based retrieval framework deployed in a large-scale content recommendation system. KuaiFormer fundamentally redefines the retrieval process by shifting from conventional score estimation tasks (such as click-through rate estimate) to a transformer-driven Next Action Prediction paradigm. 
This shift enables more effective real-time interest acquisition and multi-interest extraction, significantly enhancing retrieval performance. 
KuaiFormer has been successfully integrated into Kuaishou App's short-video recommendation system since May 2024, serving over 400 million daily active users and resulting in a marked increase in average daily usage time of Kuaishou users.
We provide insights into both the technical and business aspects of deploying transformer in large-scale recommendation systems, addressing practical challenges encountered during industrial implementation. 
Our findings offer valuable guidance for engineers and researchers aiming to leverage transformer models to optimize large-scale content recommendation systems.
\end{abstract}




\maketitle

\section{Introduction}

The Transformer \cite{transformer} architecture has demonstrated significant success across multiple domains, with notable models such as BERT \cite{bert} and GPT \cite{gpt3,gpt4} in natural language processing (NLP), and Vision Transformers \cite{vit,swintransformer,blip} in computer vision (CV). These achievements underscore the Transformer’s remarkable capabilities in sequence modeling and parallelization. In the field of recommendation systems, Transformer-based architectures like SASRec \cite{sasrec} and Bert4Rec \cite{bert4rec} have also shown potential. However, these academic efforts often fail to address certain industrial challenges, which has limited their effectiveness in driving business success in large-scale recommendation systems, such as those at Kuaishou.

Short-video recommendation poses unique challenges that demand advanced modeling techniques. The diverse nature of short-video content and the rapid evolution of user interests necessitate real-time adaptation to accurately capture these dynamic preferences. Users typically watch hundreds of short-videos each day, expressing preferences across a wide range of interest domains, while the system actively pushes diverse content to mitigate aesthetic fatigue and avoid the "filter bubble" effect. As a result, models that rely on daily updates, such as PinnerFormer\cite{pinnerformer}, struggle to adapt to users' evolving content needs. Moreover, traditional approaches like SASRec and Bert4Rec, which compress user behavior into a single interest vector, lack the capacity to accurately capture the full spectrum of user interests reflected in these interactions.

To more effectively capture complex user interests, models like MIND \cite{mind} and ComiRec \cite{comirec} employ techniques such as capsule networks to extract multiple interest vectors from user action sequences. However, because these models do not utilize native Transformer-based architectures, they are limited in fully leveraging the advantages offered by Transformers. Furthermore, they do not address the performance overhead associated with processing long sequences, which is a significant concern in industrial applications.

To effectively implement the Transformer model within Kuaishou's large-scale short video recommendation system, this study conducts a detailed analysis of these specific challenges in the recommendation field and proposes a series of concise and effective solutions tailored to address these issues.
\begin{itemize}[leftmargin=*,align=left]
\item \textbf{How to train with billion-scale item set}: 

Large language models (LLMs) typically leverage next token prediction as the pretraining task. This involves calculating the probabilities of all tokens in the vocabulary being the next token, based on the historical sequence, and selecting the one with the highest probability as the next predicted token. General-purpose language models usually contain fewer than 100,000 tokens in their vocabularies \cite{vocabularysize}. In the context of the Kuaishou short video recommendation system, the candidate pool contains billions of short videos, making it computationally prohibitive to compute probabilities for all candidates using a naive softmax approach. Efficient methods are therefore required to address the challenges of large-scale candidate selection while maintaining model performance.

\item \textbf{How to capture user's multi-interests}: Different with language which have a clear semantics direction when predicting the next token.
In our short-video services, users always have multiple interest points and a higher tolerance for short-videos watching.
As a result, there maybe exists multiple short-videos with completely different semantics that are served as `positive' next items at same time, which is unfriendly to vanilla Transformer learning.
\item \textbf{How to extend to longer sequences with fewer computation resources}: Different with large language model could stack very deeper and wider Transformers to achieve best performance with extremely higher computation resources. 
As a recommendation model, our model need to response a large amount request (about > 50 billion requests every day), thus our Retrieval model should achieve the balance between efficiency and effectiveness.
Particularly, the Transformer time complexity is $\mathcal{O}(n^2d)$, where $n$ denotes the input sequence length, $d$ means hidden state dimension \cite{keles2022computationalcomplexityselfattention}.
Thereby Transformer-based model is sensitive with sequences length and we need to devise specific module to accelerate longer sequences training.

\end{itemize}

In this work, we present KuaiFormer, our latest advancements in real-time industrial retrieval, which delivered the most significant improvements in the Retrieval stage at Kuaishou over the past year. Specifically, we introduce several reliable modifications to adapt Transformer to industrial retrieval scenarios: a customized softmax learning objective for stable model training, multiple query tokens to capture users' diverse interests, and a historical sequence compression mechanism to improve the efficiency of long-sequence modeling.
\begin{itemize}[leftmargin=*,align=left]
\item \textbf{Smooth In-Batch Softmax Loss with LogQ Correction}: To avoid directly training on a billion-scale item set, we first employ in-batch softmax as the learning objective for KuaiFormer. However, in-batch softmax inevitably introduces sampling bias, deviating from uniform item sampling: popular items have more chances to be selected as negative samples, which can lead to performance degradation. We apply the widely-used logQ correction method \cite{logq} to correct for sampling bias.
Additionally, in short-video services, users often have higher tolerance for watching, which reduces the confidence that negative samples in in-batch sampling truly represent items users dislike. Therefore, instead of using strict 0/1 labels for training, we incorporate label smoothing techniques \cite{labelsmooth} to mitigate training noise and enhance model robustness.
\item \textbf{Multi-interests Query Tokens}: To capture users' diverse interests, we drew inspiration from the [CLS] token in BERT, which introduces a learnable token to compress the original input information into a holistic sequence representation. We extend this concept in KuaiFormer by introducing multiple learnable tokens, combined with a multi-interest training strategy to extract distinct user interest representations from historical item sequences. Specifically, KuaiFormer's learnable query tokens leverage a causal masking mechanism, enabling subsequent interest tokens to fully interact with preceding interest token representations, thereby achieving more effective interest disentanglement.

\item \textbf{Adaptive Item Compression Mechanism}: To address the efficiency challenges of modeling longer sequences, we made an intuitive assumption: compared with the most recently watched short videos, users' memories of earlier videos are more vague. Therefore, we can apply coarse-grained modeling to earlier items while using fine-grained modeling for the latest ones.
Based on this, we devised an adaptive item compression mechanism: first, we divide the earlier item sequences into several groups, compressing each group into a single representation to reduce the input sequence length. This series of compressed representations is then concatenated with the most recent items, forming the token sequence input for the model.
\end{itemize}

The main contributions of our work are as follows:
\begin{itemize}[leftmargin=*,align=left]
\item We propose our next-generation retrieval approach, KuaiFormer, to our knowledge, this work is the first real-time retrieval model by the pure Transformer architecture in industrial-scale RecSys.
\item We conduct extensive offline and online experiments to show KuaiFormer superiors, which contribute  +0.360\%/+0.126\%/+0.411\% online video watch time gains to Kuaishou short-video services.
\item We offer insights into both model and architectural aspects, addressing practical challenges encountered in industrial RecSys deployment.
Our experience offer valuable guidance for engineers and researchers aiming to leverage Transformer to build better industrial RecSys.
\end{itemize}

\section{Methodology}
This section describes how KuaiFormer works in our recommendation system.
We first briefly explain the problem definition of our KuaiFormer in training and inference.
Then, we present KuaiFormer base workflow including the feature engineering and model architecture.
Next, we show the details how to model longer sequence and capture user's multiple interests.
Finally, we give our loss function to achieve a stable training in billion-scale item set.

\subsection{Problem Statement}
In a general recommendation task, we leverage users' historical watched items to model their interests and predict the next video they are likely to engage with.

For each user, let $\{(x_1,\mathrm{f}_1), (x_2,\mathrm{f}_2), \dots, (x_n,\mathrm{f}_n)\}$ represents his/her recent positive watched short-video information in chronological order, where the $x_i \in \mathbb{R}$ denotes to a short-video ID from entire short-video set $\mathcal{X}$, the $\mathrm{f}_i\in \mathbb{R}^{|\mathcal{F}|}$ denotes the watching side-information according to attribute set $\mathcal{F}$.
Generally, we hand-craft about 10+ short-video watching attributes for interests modeling, including: watching time, interaction labels (e.g., like, comment), short-video duration, multi-modal category tag ID and so on.

According to the sequence $\{(x_1,\mathrm{f}_1), (x_2,\mathrm{f}_2), \dots, (x_n,\mathrm{f}_n)\}$, in training procedure, since we know the next item $x_{n+1}$, thus we could maximize the following likelihood to capture users' real-time interests:
\begin{equation}
\begin{split}
 \{\mathbf{u}_1, \dots,& \mathbf{u}_k\} = \texttt{model}(\{(x_1,\mathrm{f}_1), (x_2,\mathrm{f}_2), \dots, (x_n,\mathrm{f}_n)\}),\\
&\texttt{argmax}_{x_{n+1} \in \mathcal{X}} \mathrm{P} \left(x_{n+1}|\{\mathbf{u}_1, \dots, \mathbf{u}_k\}\right)
\end{split}
\label{definition}
\end{equation}
where $\mathrm{P}(\cdot)\in \mathbb{R}^{|\mathcal{X}|}$ is the probability of the candidate item in $\mathcal{X}$ and $\mathbf{u}$ denotes the user interest representation.

In inference, as a retrieval model, our goal is to find a small group of candidate items that related with user's real-time interests:
\begin{equation}
\begin{split}
\{\hat{x}_{n+1}^1, \hat{x}_{n+1}^2, \dots\} = \texttt{ANN}(\{\mathbf{u}_1, \dots,& \mathbf{u}_k\}, \mathcal{X}),\\
\end{split}
\label{infer}
\end{equation}
where $\texttt{ANN}(\cdot)$ denote the Approximate Nearest Neighbor technique, which is a fast search system (e.g., faiss \cite{faiss}) to find the candidates with user interest representation $\mathbf{u}$.
The $\{\hat{x}_{n+1}^1, \hat{x}_{n+1}^2, \dots\}$ means the highest hundreds related items.

\begin{figure*}[t]
  \centering
  \includegraphics[width=1.0\textwidth]
  {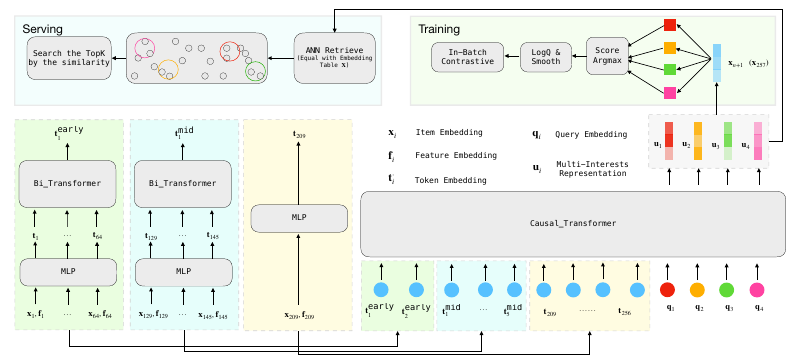}
  \caption{KuaiFormer architecture under the length setting 256, and 4 query tokens, where the $\texttt{t}^{\texttt{early}}_{1}$ denotes the early item compression, the $\texttt{t}^{\texttt{mid}}_{1}$ denotes the middle item compression. We can effectively model a longer sequence of 256 through the use of feeding a shorter sequence of 55 for efficient training and inference.}
  % \Description{blabla.}
  \label{fig:model}
\end{figure*}

\subsection{Base Backbone}

In this section, we introduce the base backbone of our KuaiFormer, which contains short-video embedding module and a stacked Transformer module.

\subsubsection{Embedding Module}
In embedding mapping stage, we utilize two type embedding layers to consider different type of attributes to form as model input:
\begin{itemize}[leftmargin=*,align=left]
\item \textit{One-hot discrete attributes embedding layer}: As the most common type of embedding layer, it is mainly used to model \textbf{discrete features, including short-video ID, tag ID, interaction labels, etc}.
Take the short-video ID mapping stage as example, we assign a parameter matrices $\mathbf{X} \in \mathbb{R}^{|\mathcal{X}|\times d}$ to store their embeddings, where $d$ is the embedding dimension.
Thereby we can easily obtain the corresponding embedding $\mathbf{x}_i \in \mathbb{R}^d$ by a straightforward lookup operation for arbitrary short-video ID $x_i$:
\begin{equation}
\begin{split}
\mathbf{x}_i = \texttt{LookUp}(\mathbf{X}, x_i),\\
\end{split}
\label{onehot}
\end{equation}
\item \textit{Bucket continuous attributes embedding layer}: Instead of the discrete attributes, this layer aims to embed the \textbf{continuous attributes, including watching time, short-video duration, etc}.
For these continuous attributes, since their prior distributions are different, we actually customize each attribute and design a special bucketing strategy to discretize them while retaining the distribution characteristics.
Take the short-video duration $\mathrm{f}^{\texttt{dura}}_i$ as example, we utilize a uniform bucket strategy to divide it in 1000 buckets with maximize 300s, and the lookup its embedding from parameter matrix $\mathbf{F}^{\texttt{duration}} \in \mathbb{R}^{1000\times d}$:
\begin{equation}
\begin{split}
\mathrm{f}^{\texttt{bucket\_dura}}_i = &\texttt{int}\Big(\frac{\texttt{min}(\mathrm{f}^{\texttt{dura}}_i, 300)}{300} * 1000\Big)\\
\mathbf{f}^{\texttt{dura}}_i = &\texttt{LookUp}(\mathbf{F}^{\texttt{dura}}, \mathrm{f}^{\texttt{bucket\_dura}}_i)\\
\end{split}
\label{bucket}
\end{equation}
where the $\mathrm{f}^{\texttt{bucket\_dura}}_i$ is a integer denotes the mapping discrete ID index for continuous attribute $\mathrm{f}^{\texttt{dura}}_i$.

\end{itemize}

\subsubsection{Transformer Module}
On top of the above embedding module, we first utilize them to form the watched short-video sequence information.
For each information $(x_i, \mathrm{f}_i)$, we can map them as:
\begin{equation}
\begin{split}
\mathbf{t}_i = \texttt{MLP}([\mathbf{x}_i, \mathbf{f}_i])\\
\end{split}
\label{token_emb}
\end{equation}
where the $\texttt{MLP}(\cdot)\in \mathbb{R}^{d\times d}$ is a feed-forward neural network, $[\cdot,\cdot]$ indicates concat operator, and the $\mathbf{t}_i$ is the final unit token representation.
By analogy, we could transform the input representation sequences as $\{\mathbf{t}_1, \mathbf{t}_2, \dots, \mathbf{t}_n\}$.

Afterwards, we utilize Transformer modeling the input sequence, to extract user interest representation.
Specifically, we follow the Llama Transformer architecture \cite{touvron2023llama} as our backbone, which mainly includes: 
(1) RMS Norm technique to avoid the gradient vanishing/explored problems,
(2) multi-head masked self-attention mechanism captures the complex token dependency, 
(3) a point-wise feed-forward layer enhances model's non-linearity capacity.
\begin{equation}
\begin{split}
\mathbf{u} = \texttt{Causal\_Transformer}(\{\mathbf{t}_1, \mathbf{t}_2, \dots, \mathbf{t}_n\}, L, M)\\
\end{split}
\label{backbone}
\end{equation}
As shown in Figure, $\mathbf{u}$ is the top layer lastest position output, $L$ is a hyper-parameter to control the depth of Transformers, $M$ is the number of multi-heads.

\subsection{Towards Longer Sequence}
%远端聚合
Actually, the generated representation $\mathbf{u}$ could support the training stage (in Eq.\ref{definition}) and inference stage (in Eq.\ref{infer}).

However, we need to handle a large volume of requests within just a few milliseconds. At peak times, this means managing over 100,000 requests per second. Thus the straightforward way is hard to scale to a longer input sequence to provide a high-quality representation for training and inference.
In our experiments, we find when the sequence length $n$ is extended from 64 to 256, the corresponding computing resources increases by 6 times, which is unacceptable.
Therefore, a challenging problem is how to model a longer sequences with fewer computation resources?
For users, the most recent videos leave a stronger impression, while older browsing history is relatively vague.
This observation motivates us to compress the earlier watched short-videos to reduce the sequence length.
In KuaiFormer, we devise a simple-yet-effective adaptive item compression mechanism in three steps:
\begin{enumerate}[leftmargin=*,align=left]
\item We first divide the input sequence into three parts according to their position as earlier part, middle part, and latest part. 
For the earlier part and middle part, we merge 64 adjacent items and 16 adjacent items as one group, respectively. 
\item For the earlier/middle item groups, we then utilize a single-layer bidirectional Transformer without mask strategy to aggregate them as a grouped item representation.
Take the earlier item group as example, such process can be formulated as:
\begin{equation}
\begin{split}
\mathbf{t}^{\texttt{early}}_1 &= \texttt{Mean}(\texttt{Bi\_Transformer}(\{\mathbf{t}_1, \dots, \mathbf{t}_{64}\}, M)\\
\mathbf{t}^{\texttt{early}}_2 &= \texttt{Mean}(\texttt{Bi\_Transformer}(\{\mathbf{t}_{65}, \dots, \mathbf{t}_{128}\}, M)\\
\mathbf{t}^{\texttt{mid}}_1 &= \texttt{Mean}(\texttt{Bi\_Transformer}(\{\mathbf{t}_{129}, \dots, \mathbf{t}_{145}\}, M)\\
&\dots \\
\mathbf{t}^{\texttt{mid}}_5 &= \texttt{Mean}(\texttt{Bi\_Transformer}(\{\mathbf{t}_{193}, \dots, \mathbf{t}_{208}\}, M)\\
\end{split}
\label{group_agg}
\end{equation}
Similarly, the calculation also holds for middle item groups, and leads to the results $\mathbf{t}^{\texttt{mid}}_i$.
\item Finally, we could explicit the compressed item sequences to replace the original input to generate user interests as:
\begin{equation}
\begin{split}
\mathbf{u} = \texttt{Causal\_Transformer}(&\{\mathbf{t}^{\texttt{early}}_1, \mathbf{t}^{\texttt{early}}_2\} \\
&\oplus\{\mathbf{t}^{\texttt{mid}}_1, \dots, \mathbf{t}^{\texttt{mid}}_5\} \\
&\oplus\{\mathbf{t}_{209}, \dots, \mathbf{t}_{256}\}, L, M) \\
\end{split}
\label{compressed_backbone}
\end{equation}
\end{enumerate}
Based on the adaptive item compression strategy, scaling the sequence length from 64 to 256, we find our model only increase 10\% additional computation resources.

\subsection{Towards Multiple Interests}
Additionally, in our short-video services, users always have multiple interest points, therefore utilizing a single representation is hard to express users' complex real-time dynamic interests.
For sequence modeling, capturing the diverse interests embedded within the sequence becomes increasingly important as the sequence length grows.
With KuaiFormer, we also explore a similar challenge: how can we model users' multi-interests within the Transformer paradigm?

Motivates by the great success of special `[CLS]' token of BERT, which introduces a learnable token to summarize input sequence information for downstream tasks.
We consider employ $k$ different special tokens act as queries token to extract different user interests.
\begin{equation}
\begin{split}
\{\mathbf{u}_1, \dots, \mathbf{u}_k\} = \texttt{Causal\_Transformer}(&\{\mathbf{t}^{\texttt{early}}_1, \mathbf{t}^{\texttt{early}}_2\} \\
&\oplus\{\mathbf{t}^{\texttt{mid}}_1, \dots, \mathbf{t}^{\texttt{mid}}_5\} \\
&\oplus\{\mathbf{t}_{209}, \dots, \mathbf{t}_{256}\} \\
&\oplus\{\mathbf{q}_1, \dots, \mathbf{q}_k\}, L, M) \\
\end{split}
\label{multi_interests}
\end{equation}
where the $\{\mathbf{q}_1, \dots, \mathbf{q}_k\}$ are $k$ different user interests query tokens, and they are concatenated with the item compression sequence.
And the $\{\mathbf{u}_1, \dots, \mathbf{u}_k\}$ are final user interests representations to fit user dynamic behaviours.
It is worth noting that in the previous retrieval approaches, their multiple interest representations are separate learned and invisible to each other.
At KuaiFormer, our query tokens also follow the auto-regressive paradigm, which allows the latter interests tokens can fully interact with former interests tokens representations to achieve better interests disentanglement.

Under the multiple interests representations, we calculate the prediction score utilize the \texttt{argmax} function.
For instance, given the next positive short-video $x_{n+1}$, we formulate the process as:
\begin{equation}
\begin{split}
\mathrm{P} (x_{n+1}|\{\mathbf{u}_1,& \dots, \mathbf{u}_k\} = \frac{\texttt{Score}_{x_{n+1}}}{\sum_{x}^{\mathcal{X}}\texttt{Score}_{x}}, \quad \texttt{where}\\
\texttt{Score}_{x_{n+1}} &= \texttt{argmax}(\{\mathbf{x}_{n+1}^\top\mathbf{u}_1, \dots, \mathbf{x}_{n+1}^\top\mathbf{u}_k\})\\
\end{split}
\label{prediction_score}
\end{equation}
where the prediction score is selected from the highest interest representation.

\begin{figure*}[t]
  \centering
  \includegraphics[width=0.85\textwidth]
  {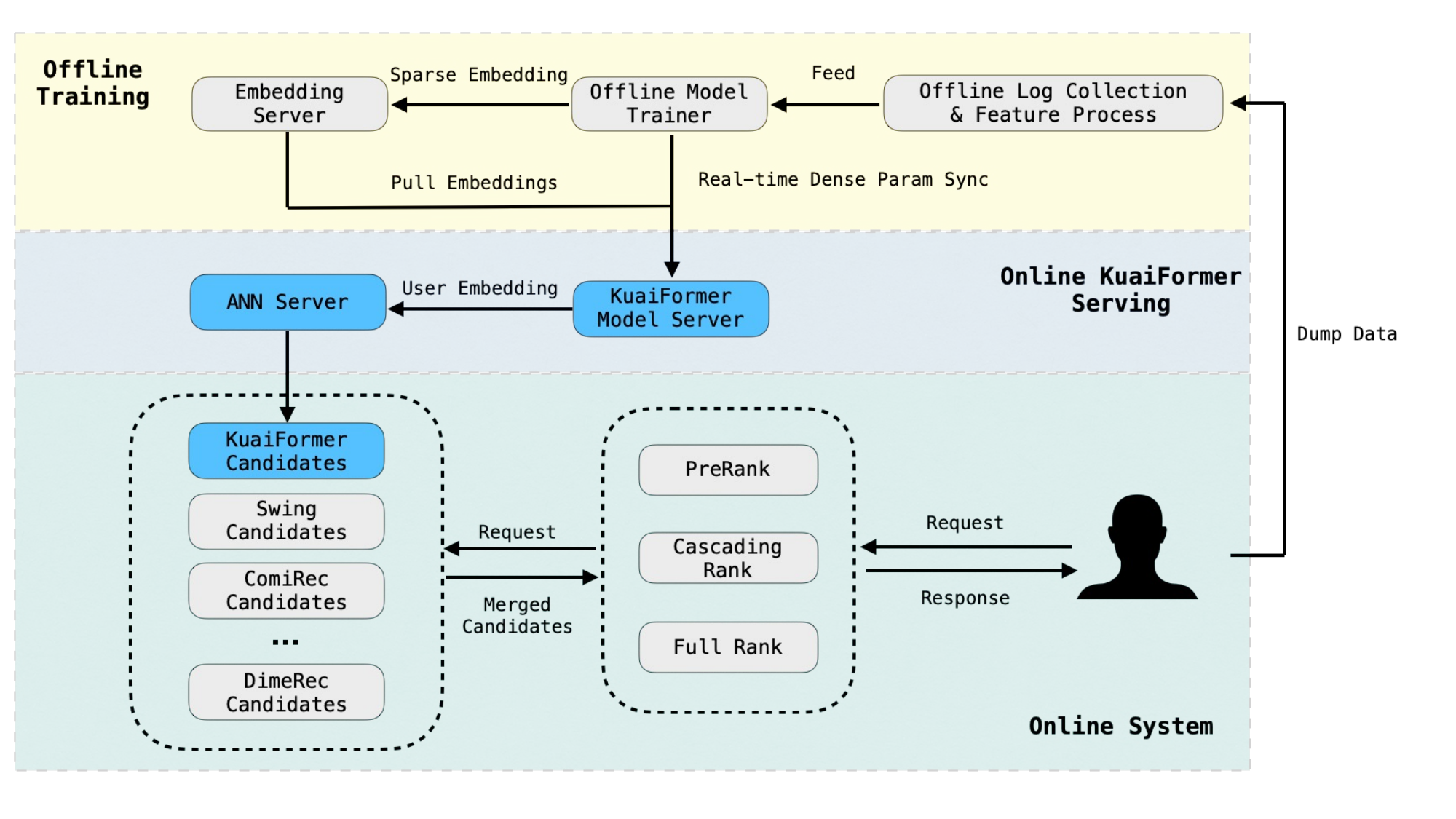}
  \caption{Deployment Architecture}
  % \Description{blabla.}
  \label{fig:serving}
\end{figure*}

\subsection{Towards Stable Training}
Up to now, we have introduced the learning process to generate multiple user interests representations with a longer sequence.
In this section, we will dive into the loss function designing to answer the most challenging problem: How to train a Transformer-based model with billion-scale item set.

Generally, the Transformer-based model always trained with softmax function, to maximize the next positive token probability while minimize all other token probabilities.
However, in recommendation task, it is impractical to train our model in full billion-scale item directly, thus we first employ the in-batch softmax as our KuaiFormer learning objective.
\begin{equation}
\begin{split}
\mathcal{L}(x_{n+1}) = -\log\Big(\frac{e^{\texttt{Score}_{x_{n+1}}}}{e^{\texttt{Score}_{x_{n+1}}} + \sum^\mathcal{B}_{\bar{x}_{n+1 }}{e^{\texttt{Score}_{\bar{x}_{n+1}}}}} \Big)
\end{split}
\label{in_batch_sample}
\end{equation}
where the $\mathcal{B}$ denotes the in-batch negative short-video set, the $x_{n+1}$ denotes the positive item and the $\bar{x}_{n+1}$ indicates an arbitrary negative item.

However, the in-batch softmax will inevitably introduce the sampling
bias to away from uniform item sampling: the top popular items
have more opportunities to become negative samples, which will
lead model performance degeneration inevitably. Therefore, we
follow the wide-used logQ correction method to refine items
sampling probability to achieve a balanced in-batch softmax \cite{logq}:
{\scriptsize
\begin{equation*}
\begin{split}
\mathcal{L}_{\texttt{logQ}} &= -\log(\texttt{Score}_{x_{n+1}}^{\texttt{logQ}}) \\
&= -\log \left( 
    \frac{
        e^{\texttt{Score}_{x_{n+1}} - \log (Q_u(x_{n+1}))}
    }{
        e^{\texttt{Score}_{x_{n+1}} - \log (Q_u(x_{n+1}))} 
        + \displaystyle\sum_{\bar{x}_{n+1} \in \mathcal{B}} 
        e^{\texttt{Score}_{\bar{x}_{n+1}} - \log (Q_u(\bar{x}_{n+1}))}
    } 
\right)
\end{split}
\label{in_batch_sample}
\end{equation*}
}

In short-video services, users tend to have a higher tolerance for watching various content, making it challenging to classify the next item as definitively "positive" or "negative." Consequently, we avoid using strict binary (0/1) labels for model training. Instead, we employ a smoothing technique to mitigate training noise and improve reliability in content labeling \cite{labelsmooth}.
\begin{equation}
	\begin{split}
		\mathcal{L}^{\texttt{smooth}}_{\texttt{logQ}} = \left\{
\begin{aligned}
-(1-\alpha)\log(\texttt{Score}_{x_{n+1}}^{\texttt{logQ}}) \\
\sum^\mathcal{B}_{\bar{x}_{n+1}}-\frac{\alpha}{|\mathcal{B}|}\log(\texttt{Score}_{\bar{x}_{n+1}}^{\texttt{logQ}}) \\
\end{aligned}
\right.
	\end{split} 
	\label{label_smooth}
\end{equation}
where the $\mathcal{L}^{\texttt{smooth}}_{\texttt{logQ}}$ is the final training objective in our KuaiFormer, $\alpha$ is a hyperparameter used to control the smoothing effect.
After our model training convergence, we utilize the ANN technique to search the top-K short-videos for each user recommendation request in Eq.(\ref{infer}).
The overall training and serving procedure are shown in Figure~\ref{fig:model}.

\begin{table*}[th]
\centering
\caption{Offline Performance (\%) comparison with increase from second-highest to highest.}
\setlength{\tabcolsep}{10pt}{
\begin{tabular}{c|cccccccc}
\toprule
Metrics & GPRP & SASRec & DimeRec & ComiRec  & Swing & GNN & KuaiFormer & Improve over runner-up \\ 
\midrule
HR@50 & 3.57\% & 1.17\% & 1.54\% & 0.95\%  & 0.14\% & \underline{3.97\%} & \textbf{4.21\%} & 6.05\% \\
\midrule
HR@100 & \underline{5.21\%} & 2.51\% & 2.49\% & 1.93\% & 0.29\% & 4.6\% & \textbf{6.58\%} & 26.30\% \\
\midrule
HR@500 & \underline{11.31\%} & 5.37\% & 6.99\% & 7.02\%  & 1.54\% & 8.89\% & \textbf{15.61\%} & 38.12\% \\
\midrule
HR@1000 & \underline{15.70\%} & 7.22\% & 9.17\% & 10.06\% & 2.03\% & 10.26\% & \textbf{19.88\%} & 26.62\% \\
\bottomrule
\end{tabular}
}
\label{mainoffline}
\end{table*}

\begin{table*}[ht]
% \footnotesize
\centering
\setlength{\tabcolsep}{5pt}
\caption{Online A/B testing results of Short-Video services at Kuaishou.}
\begin{tabular}{cccccccccc}
\toprule
\multirow{2}{*}{Applications} & \multirow{2}{*}{\makecell{Video \\Watch Time}} & \multirow{2}{*}{\makecell{Total App\\Usage Time}} & \multirow{2}{*}{\makecell{Usage Time\\per User}} & \multirow{2}{*}{\makecell{Avg. Time per\\Video View}} & \multirow{2}{*}{\makecell{Video\\Views}} & \multirow{2}{*}{\makecell{Likes}} & \multirow{2}{*}{\makecell{Follows}} & \multirow{2}{*}{\makecell{Comments}} & \multirow{2}{*}{\makecell{Novel Surprise}} \\
\\
\midrule
\makecell{Kuaishou\\Single Page}& +0.360\% & +0.184\% & +0.157\% & +0.265\% & +0.095\% & +0.256\% & +0.600\% & +0.553\% & +0.235\% \\
\midrule
\makecell{Kuaishou Lite\\Single Page} & +0.126\% & +0.127\% & +0.107\%  & +0.060\% & +0.158\% & +0.286\% & +0.222\% & +0.062\% & +0.181\% \\
\midrule
\makecell{Kuaishou\\Double Page}  & +0.411\% & -- & -- & -- & -- & +0.399\% & +1.254\% & +1.463\% & -- \\
\bottomrule
\end{tabular}

\label{mainonline}
\end{table*}

\section{Deployment Architecture}

In this section, we will present the comprehensive deployment architecture of KuaiFormer in an industrial streaming video recommender system, which serves the largest Kuaishou's recomenndation scenario in the retrieval stage. KuaiFormer continuously receives online data for training and updates its parameters to the online model service at a minute-level frequency.

As shown in Figure \ref{fig:serving}, the KuaiFormer retrieval model is trained on an industrial curated distributed training framework in an online learning way. The short video content recommendation system responds to requests from hundreds of millions of users daily. User requests first pass through the retrieval system, which is composed of multiple independent retrieval pathways, such as the classic Swing\cite{swing}, GNN\cite{lightgcn}, Comirec\cite{comirec}, Dimerec\cite{dimerec}, GPRP \cite{zheng2024full}, etc. KuaiFormer is introduced as a new retrieval pathway into the retrieval system. The retrieval candidates from all retrieval pathways are aggregated and deduplicated before being sent to the ranking system. In our system, the initial coarse-grained ranking is performed through a pre-rank stage, then a cascading rank stage, followed by a fine-grained full rank stage to obtain the final set of short videos presented to the user. The offline logging system records real-time user feedback, including dense feedback signals such as watch time and relatively sparse interactions (likes, follows, shares). All offline logs are processed into training records and aggregated for transmission to the offline training framework. To enhance system efficiency, we use a dedicated embedding server to store sparse embeddings, while the fewer dense model parameters are periodically transmitted to the online system for real-time inference of user embeddings. The final retrieval results are obtained through efficient ANN retrieval algorithms (such as Faiss\cite{faiss}, ScaNN\cite{scann}). In our practice, we employ GPU brute-force methods to retrieval the TopK candidates.

\section{Experiments}
In this section, we give detailed analyses to answer the following major research questions (RQs):
\begin{enumerate}[leftmargin=*,align=left]
    \item \textbf{RQ1}:  Does our KuaiFormer achieves SOTA offline performance compared with strong retrieval methods?
    \item \textbf{RQ2}: Does our KuaiFormer contributes online gains in our Short-Video service significantly?
    \item \textbf{RQ3}: How does different hyper-parameter settings influence KuaiFormer performance?
    % \item \textbf{RQ4}: Does KuaiFormer multi-interest work as expected?
\end{enumerate}

\subsection{Experimental Setting}

\subsubsection{Dataset}
We conduct experiments at our short-video data-streaming, which is the \textbf{largest} recommendation scenario at Kuaishou, including over \textbf{400 Million users and 50 Billion logs} every day.

\subsubsection{Evaluation Protocol}
For the performance estimation, we apply two metrics to compare with baselines and our model variants, the online Hit Rate and offline Accuracy.
\begin{itemize}[leftmargin=*,align=left]
    \item Online \textbf{Hit Rate}: For a fair comparison with different methods, we utilize the users' online requests results to estimate different model ability. Specifically, we calculate the hit rate between the real user viewed items whether they are retrieved by the corresponding models, e.g., does the viewed item is retrieved in the top 50/100/500/100 results.
    \item Offline \textbf{Accuracy}: For evaluating the performance of our KuaiFormer variants, since the models are trained in an in-batch setting and use online learning to continuously incorporate the latest logs, the accuracy when the training loss stabilizes with the same batch size can assess the model's capability. This offline training accuracy reflects the model's ability to fit user behavior.
\end{itemize}

\subsubsection{Baselines}
In Kuaishou's short video recommendation scenario, multiple retrieval models are employed simultaneously to maximize the diversity of content supply and meet user needs. We compared KuaiFormer with the following representative strong retrieval methods which are deployed at Kuaishou:
(1) Item2Item based methods: Swing \cite{swing}; (2) Multi-interests based methods: ComiRec \cite{comirec}; (3) Graph based methods: GNN \cite{lightgcn}; (4) Diffusion based methods: DimeRec \cite{dimerec}; (5) List-wise based methods: GPRP \cite{zheng2024full}.
Except them, we also implement the SASRec for a fair comparison, although it has not achieved online contribution at Kuaishou \cite{comirec}.

\begin{figure*}[h]
    \centering
    \includegraphics[width=18cm,height=5cm]{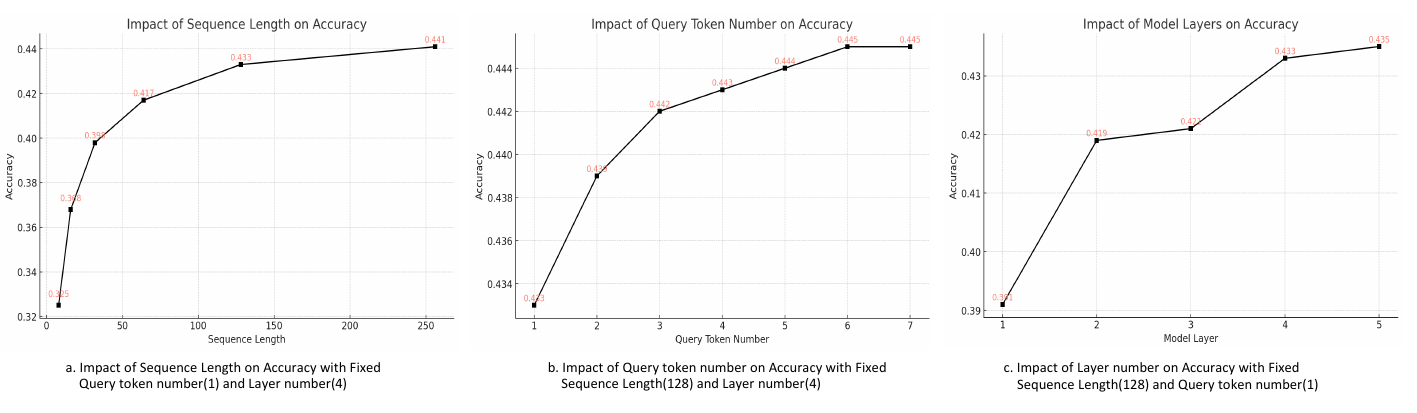}
    \caption{Impact of Sequence Length, Query token number and Layer number on Accuracy  }
    \label{fig:hyper_impact_accuracy}
\end{figure*}

\subsection{Performance Comparisons (RQ1)}
To accurately assess the offline coverage rate of the retrieval model, we replay real online requests and invoke these models, each returning 50-1000 results. 
The Hit rate between the model recommendations and the actual content viewed by users is shown in Table~\ref{mainoffline}.
According it, we have the following observations:
(1) The classic sequence modeling approach, SASRec, performs poorly in terms of hit rate, suggesting that the multi-interest module in KuaiFormer contributes to covering a broader range of user interests.
(2) Compared with the famous multi-interests method ComiRec~\cite{comirec}, our KuaiFormer shows consistent improvement in all metrics, which validates that utilizing the Transformer as base encoder to extract user preference is effective.
(3) Since GPRP is trained using recommendation funnel data from retrieval to ranking stages, it achieves a higher hit rate than other traditional methods with a retrieval number range of 100 to 1000. However, KuaiFormer improves hit rate by over 25\% compared to GPRP. This indicates that the Next Action Prediction modeling approach provides a stronger fit to the training set, even without explicitly incorporating recommendation funnel data.

\subsection{Online A/B Test (RQ2)}
To measure the precise improvements of KuaiFormer contribution to our short-video service, we conduct comprehensive one week A/B testing with 10\% of users in the three largest scenarios on Kuaishou: Kuaishou Single/Double page and Kuaishou Lite Edition Single page. In Table~\ref{mainonline}, we shown the watching-time and interaction metrics of our KuaiFormer.
Due to varying priorities across different scenarios, certain metrics are omitted on the Double Page. However, the most critical video watch time metric has been preserved to ensure essential insights remain accessible.
Notably, the about 0.1\% improvement in Video Watch Time is a statistically significant change to give satisfactory contribution, since our platform has 400 Million active users per day.
According Table~\ref{mainonline}, our KuaiFormer achieves +0.360\%, +0.126\%, +0.411\% improvements in terms of the video watch time under the three largest scenarios, \textbf{which is one of the most significant retrieval experiments at Kuaishou in 2024}. KuaiFormer has also demonstrated significant improvements in engagement metrics, such as increased Likes, Follows, and Comments. This indicates that KuaiFormer effectively adapts to evolving user needs, accurately captures user interests, and recommends relevant content, thereby enhancing user engagement and satisfaction. Additionally, the Novel Surprise metric has improved, which measures the model's ability to help users discover new interests. This suggests that the multi-interest mechanism can capture both mainstream and niche interests, introducing users to new areas of interest.

\subsection{Hyperparameter Impact (RQ3)}
This section explores four hyper-parameter effects in KuaiFomer: the sequence length, query token number, layer number and item compression mechanism.

\subsubsection{Sequence Length Impact}

As seen in the table \ref{fig:hyper_impact_accuracy} (a), increasing the sequence length from 8 to 256 steadily improves the accuracy from 0.325 to 0.441. This suggests that increasing the sequence length allows the model to capture more context, which improves its performance. However, the improvement rate decreases as the sequence length increases, indicating diminishing returns. While accuracy continues to increase, the gain from doubling the sequence length becomes smaller, especially for sequence lengths beyond 64.

\subsubsection{Query token number's impact}

The data in Table \ref{fig:hyper_impact_accuracy} (b) reveals a clear relationship between the number of query tokens and the model's accuracy. As the number of tokens increases, the accuracy improves consistently, suggesting that additional tokens provide the model with more context, enabling better predictions. However, the rate of improvement decreases as the token number rises, indicating diminishing returns. The gain in accuracy becomes smaller with each additional token, implying that the model may reach a saturation point beyond which further tokens contribute little to overall performance. For instance, the increase in accuracy from token 1 to token 2 is 0.006, while the increase from token 6 to token 7 is only 0.001, demonstrating that after a certain number of tokens, the benefit of adding more becomes minimal.

Since we retrieve a certain number of videos for each query token and then uniformly rank and truncate them to a fixed number, an excessive number of query tokens will result in fewer videos being retrieved per token. Additionally, too many query tokens will increase computational resource consumption, so the number of query tokens cannot be increased indefinitely. In this experiment, 6 tokens appear to represent an optimal balance, where accuracy gains begin to level off, making it unnecessary to further increase the number of tokens for substantial improvements.

\subsubsection{Layer number's impact}

Table \ref{fig:hyper_impact_accuracy} (c) demonstrates the impact of increasing model layers on accuracy, while holding the sequence length and query token number constant. The results show a consistent improvement in accuracy as the number of layers increases from 1 to 5.

The increase in accuracy from Layer 1 (0.391) to Layer 2 (0.419) represents a substantial gain, suggesting that additional layers contribute to more refined feature extraction and better model performance. However, similar to the effect observed with query tokens, the improvement diminishes with each additional layer. For instance, the increase from Layer 4 (0.433) to Layer 5 (0.435) is only 0.002, indicating that after a certain depth, the model reaches a point of diminishing returns, where adding more layers provides marginal benefits.

This pattern suggests that while deeper models generally perform better, there is a trade-off between complexity and performance. Beyond a certain depth (around Layer 4 or 5 in this case), the accuracy gains become marginal, and additional layers may not justify the increased computational cost. Thus, in this context, Layer 4 or 5 appears to be the optimal depth, balancing accuracy with efficiency.

\subsubsection{Item compression strategy impact}

\begin{table}[t]
    \centering
    \resizebox{\linewidth}{!}{
    \begin{tabular}{l|ccc}
        \toprule
        \multirow{2}{*}{\makecell{Modifications}} & \multicolumn{3}{c}{Accuracy} \\
        \cmidrule(r){2-4} & Variant1 & Variant2 & Variant3 \\
        \hline
        Original Sequence Length &64 &256 &256 \\
        Compressed Sequence Length &\ding{56} &\ding{56} &\ding{52} \\
        \hline
        Accuracy &0.417 &0.441 &0.445 \\
        \bottomrule
    \end{tabular}
    }
    \caption{Comparison of Model Performance with Different Sequence Lengths and Compression}
    \label{tab:compression_comparison}
\end{table}

Table \ref{tab:compression_comparison} compares the performance of models with varying sequence lengths and compression strategies. Specifically, we explored the impact of compressing item sequences on model accuracy.

For the compression approach, we devised a method where a sequence of 256 items is progressively compressed with compression window sizes of 64, 64, 16, 16, 16, 16, and 16 items, starting from the oldest to the most recent. The latest 48 items are appended to this sequence without compression, and the resulting sequence is then fed into the model. The query token count is fixed at 1, and the model architecture comprises a 4-layer transformer.

Our findings indicate that the compressed sequence outperforms the model with a sequence length of 64 and, notably, achieves slightly better accuracy than the model with the full uncompressed sequence of 256 items. This suggests that the compression strategy effectively reduces noise in the sequence, contributing to improved model performance.

\section{Related Works}
\subsection{Sequential Recommendation}

In recommender systems, the recommendation process is typically divided into two key stages: retrieval and ranking. The retrieval stage aims to efficiently retrieve a subset of items from a large candidate pool that a user might be interested in. Due to computational constraints, retrieval models are often designed with lighter architectures, frequently employing dual-tower structures where separate towers model users and items, interacting only at the top layer \cite{ebr,youtubednn}. In contrast, the ranking stage scores the retrieved subset, allowing for the application of more complex models and intricate feature interactions \cite{zhou2018deep,zhou2019deep,chang2023pepnet,pi2019practice}. The focus of this work is to design more effective retrieval models.

Historically, collaborative filtering methods have been widely used in recommendation tasks, including item-based CF, user-based CF, and factorization machines \cite{herlocker2000explaining,koren2021advances,ko2022survey}. With advancements in deep learning, models such as DNN and Embedding-Based Retrieval have become prevalent \cite{youtubednn,ebr,li2021embedding,huang2013learning}. In the realm of recommendation, leveraging users' historical behavior is crucial. Predicting the next item a user might engage with based on their behavior sequence is known as sequential modeling \cite{smirnova2017contextual,liu2016context,wang2019sequential}.

Several works have employed Transformer architectures for sequential modeling. For instance, SASRec models the recommendation task as an autoregressive problem, predicting the next item in a sequence based on previous interactions \cite{sasrec}. In contrast, BERT4Rec utilizes bidirectional self-attention mechanisms to capture both past and future contexts within a sequence, enhancing the model's understanding of user behavior \cite{bert4rec}. PinnerFormer, implemented in real-world industrial scenarios, leverages Transformer architectures to capture users' long-term interests for recommendation purposes \cite{pinnerformer}.

These Transformer-based models have demonstrated significant improvements in capturing complex user behavior patterns, thereby enhancing the effectiveness of recommender systems. 

\subsection{Multi-Interest User Representation}
In recommendation systems, users often exhibit diverse interests, necessitating the provision of varied content to meet their multifaceted preferences. Traditional models typically represent a user's interest with a single vector, which may not adequately capture the complexity of user behavior. To address this limitation, several approaches have been developed to model multiple user interests, particularly during the retrieval phase.
The MIND model assigns a fixed number of vectors to each user, representing preferences for different content types \cite{mind}. It employs a dynamic routing mechanism from capsule networks to extract multiple interest representations from user behavior sequences, thereby capturing diverse user interests more effectively. 
Building upon MIND, the ComiRec model explores methods to integrate multi-interest retrieval results \cite{comirec}. ComiRec introduces a controllable aggregation module designed to balance recommendation accuracy and diversity. Additionally, it incorporates multi-interest extraction modules based on dynamic routing and self-attention mechanisms to better capture users' multiple interests. 
The MIP utilizes a time-aware self-attention mechanism to extract multiple interest representations from user behavior sequences \cite{shi2023everyone}. By incorporating temporal information, MIP more accurately captures the dynamic evolution of user interests, thereby enhancing recommendation performance.

\section{Conclusion}
KuaiFormer demonstrates the feasibility and effectiveness of a Transformer-based architecture for large-scale retrieval tasks in a short-video recommendation system. By leveraging multi-interest extraction, adaptive sequence compression, and stable training techniques, KuaiFormer effectively captures the complex, dynamic interests of users and scales efficiently across billions of requests. Our extensive offline and online evaluations confirm significant improvements in both offline hit rate and online ab test, validating KuaiFormer's ability to enhance user satisfaction and business performance. This work underscores the transformative potential of advanced neural architectures in industrial recommendation systems, offering a scalable framework that can inspire further innovations in content retrieval and recommendation.

\balance
\bibliographystyle{ACM-Reference-Format}
\bibliography{sample-sigconf.bib}


\begin{thebibliography}{38}

%%% the Web address.

\ifx \showCODEN    \undefined \def \showCODEN     #1{\unskip}     \fi
\ifx \showDOI      \undefined \def \showDOI       #1{#1}\fi
\ifx \showISBNx    \undefined \def \showISBNx     #1{\unskip}     \fi
\ifx \showISBNxiii \undefined \def \showISBNxiii  #1{\unskip}     \fi
\ifx \showISSN     \undefined \def \showISSN      #1{\unskip}     \fi
\ifx \showLCCN     \undefined \def \showLCCN      #1{\unskip}     \fi
\ifx \shownote     \undefined \def \shownote      #1{#1}          \fi
\ifx \showarticletitle \undefined \def \showarticletitle #1{#1}   \fi
\ifx \showURL      \undefined \def \showURL       {\relax}        \fi
% invisible to TeX
\providecommand\bibfield[2]{#2}
\providecommand\bibinfo[2]{#2}
\providecommand\natexlab[1]{#1}
\providecommand\showeprint[2][]{arXiv:#2}

\bibitem[Achiam et~al\mbox{.}(2023)]%
        {gpt4}
\bibfield{author}{\bibinfo{person}{Josh Achiam}, \bibinfo{person}{Steven Adler}, \bibinfo{person}{Sandhini Agarwal}, \bibinfo{person}{Lama Ahmad}, \bibinfo{person}{Ilge Akkaya}, \bibinfo{person}{Florencia~Leoni Aleman}, \bibinfo{person}{Diogo Almeida}, \bibinfo{person}{Janko Altenschmidt}, \bibinfo{person}{Sam Altman}, \bibinfo{person}{Shyamal Anadkat}, {et~al\mbox{.}}} \bibinfo{year}{2023}\natexlab{}.
\newblock \showarticletitle{Gpt-4 technical report}.
\newblock


\bibitem[Alexey(2020)]%
        {vit}
\bibfield{author}{\bibinfo{person}{Dosovitskiy Alexey}.} \bibinfo{year}{2020}\natexlab{}.
\newblock \showarticletitle{An image is worth 16x16 words: Transformers for image recognition at scale}. In \bibinfo{booktitle}{\emph{arXiv}}.
\newblock


\bibitem[Brown(2020)]%
        {gpt3}
\bibfield{author}{\bibinfo{person}{Tom~B Brown}.} \bibinfo{year}{2020}\natexlab{}.
\newblock \showarticletitle{Language models are few-shot learners}. In \bibinfo{booktitle}{\emph{arXiv}}.
\newblock


\bibitem[Cen et~al\mbox{.}(2020)]%
        {comirec}
\bibfield{author}{\bibinfo{person}{Yukuo Cen}, \bibinfo{person}{Jianwei Zhang}, \bibinfo{person}{Xu Zou}, \bibinfo{person}{Chang Zhou}, \bibinfo{person}{Hongxia Yang}, {and} \bibinfo{person}{Jie Tang}.} \bibinfo{year}{2020}\natexlab{}.
\newblock \showarticletitle{Controllable multi-interest framework for recommendation}. In \bibinfo{booktitle}{\emph{ACM SIGKDD Conference on Knowledge Discovery and Data Mining (KDD)}}. \bibinfo{pages}{2942--2951}.
\newblock


\bibitem[Chang et~al\mbox{.}(2023)]%
        {chang2023pepnet}
\bibfield{author}{\bibinfo{person}{Jianxin Chang}, \bibinfo{person}{Chenbin Zhang}, \bibinfo{person}{Yiqun Hui}, \bibinfo{person}{Dewei Leng}, \bibinfo{person}{Yanan Niu}, \bibinfo{person}{Yang Song}, {and} \bibinfo{person}{Kun Gai}.} \bibinfo{year}{2023}\natexlab{}.
\newblock \showarticletitle{Pepnet: Parameter and embedding personalized network for infusing with personalized prior information}. In \bibinfo{booktitle}{\emph{ACM SIGKDD Conference on Knowledge Discovery and Data Mining (KDD)}}.
\newblock


\bibitem[Covington et~al\mbox{.}(2016)]%
        {youtubednn}
\bibfield{author}{\bibinfo{person}{Paul Covington}, \bibinfo{person}{Jay Adams}, {and} \bibinfo{person}{Emre Sargin}.} \bibinfo{year}{2016}\natexlab{}.
\newblock \showarticletitle{Deep neural networks for youtube recommendations}. In \bibinfo{booktitle}{\emph{ACM Conference on Recommender Systems (RecSys)}}.
\newblock


\bibitem[Devlin et~al\mbox{.}(2019)]%
        {bert}
\bibfield{author}{\bibinfo{person}{Jacob Devlin}, \bibinfo{person}{Ming-Wei Chang}, \bibinfo{person}{Kenton Lee}, {and} \bibinfo{person}{Kristina Toutanova}.} \bibinfo{year}{2019}\natexlab{}.
\newblock \showarticletitle{{BERT}: Pre-training of Deep Bidirectional Transformers for Language Understanding}. In \bibinfo{booktitle}{\emph{NAACL}}, \bibfield{editor}{\bibinfo{person}{Jill Burstein}, \bibinfo{person}{Christy Doran}, {and} \bibinfo{person}{Thamar Solorio}} (Eds.).
\newblock


\bibitem[Guo et~al\mbox{.}(2020)]%
        {scann}
\bibfield{author}{\bibinfo{person}{Ruiqi Guo}, \bibinfo{person}{Philip Sun}, \bibinfo{person}{Erik Lindgren}, \bibinfo{person}{Quan Geng}, \bibinfo{person}{David Simcha}, \bibinfo{person}{Felix Chern}, {and} \bibinfo{person}{Sanjiv Kumar}.} \bibinfo{year}{2020}\natexlab{}.
\newblock \showarticletitle{Accelerating large-scale inference with anisotropic vector quantization}. In \bibinfo{booktitle}{\emph{International Conference on Machine Learning}}. PMLR, \bibinfo{pages}{3887--3896}.
\newblock


\bibitem[He et~al\mbox{.}(2020)]%
        {lightgcn}
\bibfield{author}{\bibinfo{person}{Xiangnan He}, \bibinfo{person}{Kuan Deng}, \bibinfo{person}{Xiang Wang}, \bibinfo{person}{Yan Li}, \bibinfo{person}{Yongdong Zhang}, {and} \bibinfo{person}{Meng Wang}.} \bibinfo{year}{2020}\natexlab{}.
\newblock \showarticletitle{Lightgcn: Simplifying and powering graph convolution network for recommendation}. In \bibinfo{booktitle}{\emph{International ACM SIGIR Conference on Research and Development in Information Retrieval (SIGIR)}}.
\newblock


\bibitem[Herlocker et~al\mbox{.}(2000)]%
        {herlocker2000explaining}
\bibfield{author}{\bibinfo{person}{Jonathan~L Herlocker}, \bibinfo{person}{Joseph~A Konstan}, {and} \bibinfo{person}{John Riedl}.} \bibinfo{year}{2000}\natexlab{}.
\newblock \showarticletitle{Explaining collaborative filtering recommendations}. In \bibinfo{booktitle}{\emph{ACM conference on Computer supported cooperative work}}.
\newblock


\bibitem[Huang et~al\mbox{.}(2020)]%
        {ebr}
\bibfield{author}{\bibinfo{person}{Jui-Ting Huang}, \bibinfo{person}{Ashish Sharma}, \bibinfo{person}{Shuying Sun}, \bibinfo{person}{Li Xia}, \bibinfo{person}{David Zhang}, \bibinfo{person}{Philip Pronin}, \bibinfo{person}{Janani Padmanabhan}, \bibinfo{person}{Giuseppe Ottaviano}, {and} \bibinfo{person}{Linjun Yang}.} \bibinfo{year}{2020}\natexlab{}.
\newblock \showarticletitle{Embedding-based retrieval in facebook search}. In \bibinfo{booktitle}{\emph{ACM SIGKDD Conference on Knowledge Discovery and Data Mining (KDD)}}.
\newblock


\bibitem[Huang et~al\mbox{.}(2013)]%
        {huang2013learning}
\bibfield{author}{\bibinfo{person}{Po-Sen Huang}, \bibinfo{person}{Xiaodong He}, \bibinfo{person}{Jianfeng Gao}, \bibinfo{person}{Li Deng}, \bibinfo{person}{Alex Acero}, {and} \bibinfo{person}{Larry Heck}.} \bibinfo{year}{2013}\natexlab{}.
\newblock \showarticletitle{Learning deep structured semantic models for web search using clickthrough data}. In \bibinfo{booktitle}{\emph{ACM International Conference on Information and Knowledge Management (CIKM)}}.
\newblock


\bibitem[Johnson et~al\mbox{.}(2019)]%
        {faiss}
\bibfield{author}{\bibinfo{person}{Jeff Johnson}, \bibinfo{person}{Matthijs Douze}, {and} \bibinfo{person}{Herv{\'e} J{\'e}gou}.} \bibinfo{year}{2019}\natexlab{}.
\newblock \showarticletitle{Billion-scale similarity search with GPUs}. In \bibinfo{booktitle}{\emph{IEEE Transactions on Big Data}}.
\newblock


\bibitem[Kang and McAuley(2018)]%
        {sasrec}
\bibfield{author}{\bibinfo{person}{Wang-Cheng Kang} {and} \bibinfo{person}{Julian McAuley}.} \bibinfo{year}{2018}\natexlab{}.
\newblock \showarticletitle{Self-attentive sequential recommendation}. In \bibinfo{booktitle}{\emph{IEEE international conference on data mining (ICDM)}}.
\newblock


\bibitem[Keles et~al\mbox{.}(2022)]%
        {keles2022computationalcomplexityselfattention}
\bibfield{author}{\bibinfo{person}{Feyza~Duman Keles}, \bibinfo{person}{Pruthuvi~Mahesakya Wijewardena}, {and} \bibinfo{person}{Chinmay Hegde}.} \bibinfo{year}{2022}\natexlab{}.
\newblock \showarticletitle{On The Computational Complexity of Self-Attention}. In \bibinfo{booktitle}{\emph{arXiv}}.
\newblock


\bibitem[Ko et~al\mbox{.}(2022)]%
        {ko2022survey}
\bibfield{author}{\bibinfo{person}{Hyeyoung Ko}, \bibinfo{person}{Suyeon Lee}, \bibinfo{person}{Yoonseo Park}, {and} \bibinfo{person}{Anna Choi}.} \bibinfo{year}{2022}\natexlab{}.
\newblock \showarticletitle{A survey of recommendation systems: recommendation models, techniques, and application fields}.
\newblock \bibinfo{journal}{\emph{Electronics}} (\bibinfo{year}{2022}).
\newblock


\bibitem[Koren et~al\mbox{.}(2021)]%
        {koren2021advances}
\bibfield{author}{\bibinfo{person}{Yehuda Koren}, \bibinfo{person}{Steffen Rendle}, {and} \bibinfo{person}{Robert Bell}.} \bibinfo{year}{2021}\natexlab{}.
\newblock \showarticletitle{Advances in collaborative filtering}.
\newblock \bibinfo{journal}{\emph{Recommender systems handbook}} (\bibinfo{year}{2021}).
\newblock


\bibitem[Li et~al\mbox{.}(2019)]%
        {mind}
\bibfield{author}{\bibinfo{person}{Chao Li}, \bibinfo{person}{Zhiyuan Liu}, \bibinfo{person}{Mengmeng Wu}, \bibinfo{person}{Yuchi Xu}, \bibinfo{person}{Huan Zhao}, \bibinfo{person}{Pipei Huang}, \bibinfo{person}{Guoliang Kang}, \bibinfo{person}{Qiwei Chen}, \bibinfo{person}{Wei Li}, {and} \bibinfo{person}{Dik~Lun Lee}.} \bibinfo{year}{2019}\natexlab{}.
\newblock \showarticletitle{Multi-interest network with dynamic routing for recommendation at Tmall}. In \bibinfo{booktitle}{\emph{ACM International Conference on Information and Knowledge Management (CIKM)}}.
\newblock


\bibitem[Li et~al\mbox{.}(2022)]%
        {blip}
\bibfield{author}{\bibinfo{person}{Junnan Li}, \bibinfo{person}{Dongxu Li}, \bibinfo{person}{Caiming Xiong}, {and} \bibinfo{person}{Steven Hoi}.} \bibinfo{year}{2022}\natexlab{}.
\newblock \showarticletitle{Blip: Bootstrapping language-image pre-training for unified vision-language understanding and generation}. In \bibinfo{booktitle}{\emph{International conference on machine learning}}.
\newblock


\bibitem[Li et~al\mbox{.}(2021)]%
        {li2021embedding}
\bibfield{author}{\bibinfo{person}{Sen Li}, \bibinfo{person}{Fuyu Lv}, \bibinfo{person}{Taiwei Jin}, \bibinfo{person}{Guli Lin}, \bibinfo{person}{Keping Yang}, \bibinfo{person}{Xiaoyi Zeng}, \bibinfo{person}{Xiao-Ming Wu}, {and} \bibinfo{person}{Qianli Ma}.} \bibinfo{year}{2021}\natexlab{}.
\newblock \showarticletitle{Embedding-based product retrieval in taobao search}. In \bibinfo{booktitle}{\emph{ACM SIGKDD Conference on Knowledge Discovery and Data Mining (KDD)}}.
\newblock


\bibitem[Li et~al\mbox{.}(2024)]%
        {dimerec}
\bibfield{author}{\bibinfo{person}{Wuchao Li}, \bibinfo{person}{Rui Huang}, \bibinfo{person}{Haijun Zhao}, \bibinfo{person}{Chi Liu}, \bibinfo{person}{Kai Zheng}, \bibinfo{person}{Qi Liu}, \bibinfo{person}{Na Mou}, \bibinfo{person}{Guorui Zhou}, \bibinfo{person}{Defu Lian}, \bibinfo{person}{Yang Song}, {et~al\mbox{.}}} \bibinfo{year}{2024}\natexlab{}.
\newblock \showarticletitle{DimeRec: A Unified Framework for Enhanced Sequential Recommendation via Generative Diffusion Models}. In \bibinfo{booktitle}{\emph{arXiv}}.
\newblock


\bibitem[Liu et~al\mbox{.}(2016)]%
        {liu2016context}
\bibfield{author}{\bibinfo{person}{Qiang Liu}, \bibinfo{person}{Shu Wu}, \bibinfo{person}{Diyi Wang}, \bibinfo{person}{Zhaokang Li}, {and} \bibinfo{person}{Liang Wang}.} \bibinfo{year}{2016}\natexlab{}.
\newblock \showarticletitle{Context-aware sequential recommendation}. In \bibinfo{booktitle}{\emph{IEEE international conference on data mining (ICDM)}}.
\newblock


\bibitem[Liu et~al\mbox{.}(2021)]%
        {swintransformer}
\bibfield{author}{\bibinfo{person}{Ze Liu}, \bibinfo{person}{Yutong Lin}, \bibinfo{person}{Yue Cao}, \bibinfo{person}{Han Hu}, \bibinfo{person}{Yixuan Wei}, \bibinfo{person}{Zheng Zhang}, \bibinfo{person}{Stephen Lin}, {and} \bibinfo{person}{Baining Guo}.} \bibinfo{year}{2021}\natexlab{}.
\newblock \showarticletitle{Swin transformer: Hierarchical vision transformer using shifted windows}. In \bibinfo{booktitle}{\emph{Proceedings of the IEEE/CVF international conference on computer vision}}. \bibinfo{pages}{10012--10022}.
\newblock


\bibitem[M{\"{u}}ller et~al\mbox{.}(2019)]%
        {labelsmooth}
\bibfield{author}{\bibinfo{person}{Rafael M{\"{u}}ller}, \bibinfo{person}{Simon Kornblith}, {and} \bibinfo{person}{Geoffrey~E. Hinton}.} \bibinfo{year}{2019}\natexlab{}.
\newblock \showarticletitle{When Does Label Smoothing Help?}. In \bibinfo{booktitle}{\emph{arXiv}}.
\newblock


\bibitem[Pancha et~al\mbox{.}(2022)]%
        {pinnerformer}
\bibfield{author}{\bibinfo{person}{Nikil Pancha}, \bibinfo{person}{Andrew Zhai}, \bibinfo{person}{Jure Leskovec}, {and} \bibinfo{person}{Charles Rosenberg}.} \bibinfo{year}{2022}\natexlab{}.
\newblock \showarticletitle{PinnerFormer: Sequence Modeling for User Representation at Pinterest}. In \bibinfo{booktitle}{\emph{ACM SIGKDD Conference on Knowledge Discovery and Data Mining (KDD)}}.
\newblock


\bibitem[Pi et~al\mbox{.}(2019)]%
        {pi2019practice}
\bibfield{author}{\bibinfo{person}{Qi Pi}, \bibinfo{person}{Weijie Bian}, \bibinfo{person}{Guorui Zhou}, \bibinfo{person}{Xiaoqiang Zhu}, {and} \bibinfo{person}{Kun Gai}.} \bibinfo{year}{2019}\natexlab{}.
\newblock \showarticletitle{Practice on long sequential user behavior modeling for click-through rate prediction}. In \bibinfo{booktitle}{\emph{ACM SIGKDD Conference on Knowledge Discovery and Data Mining (KDD)}}.
\newblock


\bibitem[Shi et~al\mbox{.}(2023)]%
        {shi2023everyone}
\bibfield{author}{\bibinfo{person}{Hui Shi}, \bibinfo{person}{Yupeng Gu}, \bibinfo{person}{Yitong Zhou}, \bibinfo{person}{Bo Zhao}, \bibinfo{person}{Sicun Gao}, {and} \bibinfo{person}{Jishen Zhao}.} \bibinfo{year}{2023}\natexlab{}.
\newblock \showarticletitle{Everyone’s preference changes differently: A weighted multi-interest model for retrieval}. In \bibinfo{booktitle}{\emph{International Conference on Machine Learning}}. \bibinfo{pages}{31228--31242}.
\newblock


\bibitem[Smirnova and Vasile(2017)]%
        {smirnova2017contextual}
\bibfield{author}{\bibinfo{person}{Elena Smirnova} {and} \bibinfo{person}{Flavian Vasile}.} \bibinfo{year}{2017}\natexlab{}.
\newblock \showarticletitle{Contextual sequence modeling for recommendation with recurrent neural networks}. In \bibinfo{booktitle}{\emph{Proceedings of the 2nd workshop on deep learning for recommender systems}}.
\newblock


\bibitem[Sun et~al\mbox{.}(2019)]%
        {bert4rec}
\bibfield{author}{\bibinfo{person}{Fei Sun}, \bibinfo{person}{Jun Liu}, \bibinfo{person}{Jian Wu}, \bibinfo{person}{Changhua Pei}, \bibinfo{person}{Xiao Lin}, \bibinfo{person}{Wenwu Ou}, {and} \bibinfo{person}{Peng Jiang}.} \bibinfo{year}{2019}\natexlab{}.
\newblock \showarticletitle{BERT4Rec: Sequential recommendation with bidirectional encoder representations from transformer}. In \bibinfo{booktitle}{\emph{ACM International Conference on Information and Knowledge Management (CIKM)}}.
\newblock


\bibitem[Tao et~al\mbox{.}(2024)]%
        {vocabularysize}
\bibfield{author}{\bibinfo{person}{Chaofan Tao}, \bibinfo{person}{Qian Liu}, \bibinfo{person}{Longxu Dou}, \bibinfo{person}{Niklas Muennighoff}, \bibinfo{person}{Zhongwei Wan}, \bibinfo{person}{Ping Luo}, \bibinfo{person}{Min Lin}, {and} \bibinfo{person}{Ngai Wong}.} \bibinfo{year}{2024}\natexlab{}.
\newblock \showarticletitle{Scaling laws with vocabulary: Larger models deserve larger vocabularies}. In \bibinfo{booktitle}{\emph{arXiv}}.
\newblock


\bibitem[Touvron et~al\mbox{.}(2023)]%
        {touvron2023llama}
\bibfield{author}{\bibinfo{person}{Hugo Touvron}, \bibinfo{person}{Thibaut Lavril}, \bibinfo{person}{Gautier Izacard}, \bibinfo{person}{Xavier Martinet}, \bibinfo{person}{Marie-Anne Lachaux}, \bibinfo{person}{Timoth{\'e}e Lacroix}, \bibinfo{person}{Baptiste Rozi{\`e}re}, \bibinfo{person}{Naman Goyal}, \bibinfo{person}{Eric Hambro}, \bibinfo{person}{Faisal Azhar}, {et~al\mbox{.}}} \bibinfo{year}{2023}\natexlab{}.
\newblock \showarticletitle{Llama: Open and efficient foundation language models}. In \bibinfo{booktitle}{\emph{arXiv}}.
\newblock


\bibitem[Vaswani(2017)]%
        {transformer}
\bibfield{author}{\bibinfo{person}{A Vaswani}.} \bibinfo{year}{2017}\natexlab{}.
\newblock \showarticletitle{Attention is all you need}.
\newblock \bibinfo{journal}{\emph{Advances in Neural Information Processing Systems (NeurIPS)}} (\bibinfo{year}{2017}).
\newblock


\bibitem[Wang et~al\mbox{.}(2019)]%
        {wang2019sequential}
\bibfield{author}{\bibinfo{person}{Shoujin Wang}, \bibinfo{person}{Liang Hu}, \bibinfo{person}{Yan Wang}, \bibinfo{person}{Longbing Cao}, \bibinfo{person}{Quan~Z Sheng}, {and} \bibinfo{person}{Mehmet Orgun}.} \bibinfo{year}{2019}\natexlab{}.
\newblock \showarticletitle{Sequential recommender systems: challenges, progress and prospects}.
\newblock \bibinfo{journal}{\emph{arXiv}} (\bibinfo{year}{2019}).
\newblock


\bibitem[Yang et~al\mbox{.}(2020)]%
        {swing}
\bibfield{author}{\bibinfo{person}{Xiaoyong Yang}, \bibinfo{person}{Yadong Zhu}, \bibinfo{person}{Yi Zhang}, \bibinfo{person}{Xiaobo Wang}, {and} \bibinfo{person}{Quan Yuan}.} \bibinfo{year}{2020}\natexlab{}.
\newblock \showarticletitle{Large scale product graph construction for recommendation in e-commerce}. In \bibinfo{booktitle}{\emph{arXiv}}.
\newblock


\bibitem[Yi et~al\mbox{.}(2019)]%
        {logq}
\bibfield{author}{\bibinfo{person}{Xinyang Yi}, \bibinfo{person}{Ji Yang}, \bibinfo{person}{Lichan Hong}, \bibinfo{person}{Derek~Zhiyuan Cheng}, \bibinfo{person}{Lukasz Heldt}, \bibinfo{person}{Aditee~Ajit Kumthekar}, \bibinfo{person}{Zhe Zhao}, \bibinfo{person}{Li Wei}, {and} \bibinfo{person}{Ed Chi}.} \bibinfo{year}{2019}\natexlab{}.
\newblock \showarticletitle{Sampling-Bias-Corrected Neural Modeling for Large Corpus Item Recommendations}. In \bibinfo{booktitle}{\emph{ACM Conference on Recommender Systems (RecSys)}}.
\newblock


\bibitem[Zheng et~al\mbox{.}(2024)]%
        {zheng2024full}
\bibfield{author}{\bibinfo{person}{Kai Zheng}, \bibinfo{person}{Haijun Zhao}, \bibinfo{person}{Rui Huang}, \bibinfo{person}{Beichuan Zhang}, \bibinfo{person}{Na Mou}, \bibinfo{person}{Yanan Niu}, \bibinfo{person}{Yang Song}, \bibinfo{person}{Hongning Wang}, {and} \bibinfo{person}{Kun Gai}.} \bibinfo{year}{2024}\natexlab{}.
\newblock \showarticletitle{Full Stage Learning to Rank: A Unified Framework for Multi-Stage Systems}. In \bibinfo{booktitle}{\emph{Proceedings of the ACM on Web Conference}}.
\newblock


\bibitem[Zhou et~al\mbox{.}(2019)]%
        {zhou2019deep}
\bibfield{author}{\bibinfo{person}{Guorui Zhou}, \bibinfo{person}{Na Mou}, \bibinfo{person}{Ying Fan}, \bibinfo{person}{Qi Pi}, \bibinfo{person}{Weijie Bian}, \bibinfo{person}{Chang Zhou}, \bibinfo{person}{Xiaoqiang Zhu}, {and} \bibinfo{person}{Kun Gai}.} \bibinfo{year}{2019}\natexlab{}.
\newblock \showarticletitle{Deep interest evolution network for click-through rate prediction}. In \bibinfo{booktitle}{\emph{Proceedings of the AAAI conference on artificial intelligence}}.
\newblock


\bibitem[Zhou et~al\mbox{.}(2018)]%
        {zhou2018deep}
\bibfield{author}{\bibinfo{person}{Guorui Zhou}, \bibinfo{person}{Xiaoqiang Zhu}, \bibinfo{person}{Chenru Song}, \bibinfo{person}{Ying Fan}, \bibinfo{person}{Han Zhu}, \bibinfo{person}{Xiao Ma}, \bibinfo{person}{Yanghui Yan}, \bibinfo{person}{Junqi Jin}, \bibinfo{person}{Han Li}, {and} \bibinfo{person}{Kun Gai}.} \bibinfo{year}{2018}\natexlab{}.
\newblock \showarticletitle{Deep interest network for click-through rate prediction}. In \bibinfo{booktitle}{\emph{ACM SIGKDD Conference on Knowledge Discovery and Data Mining (KDD)}}.
\newblock


\end{thebibliography}
\end{document}